# Development of a High-Doppler Shift Optical Simulator to Test Femtosecond-Level Optical Time-Frequency Transfer

**Matthew S. Bigelow[1*], Kyle W. Martin[1], Nolan Matthews[2], Benjamin K. Stuhl[2], Richard Gonzales[1], Alexandre de Pinho e Braga[1], John Elgin[3], and Kimberly A. Frey[3]**

[1] *AV Inc., 1300 Britt St SE, Albuquerque, NM 87123, USA*
[2] *Space Dynamics Laboratory, North Logan, UT 84341, USA*
[3] *Air Force Research Lab, Kirtland Air Force Base, NM 87117, USA*
**matt.bigelow@bluehalo.com*

**Abstract:** We have developed an optical simulator to test optical two-way time-frequency transfer (O-TWTFT) at the femtosecond level capable of simulating relative motion between two linked optical clock nodes up to Mach 1.8 with no moving parts. The technique is enabled by artificially Doppler shifting femtosecond pulses from auxiliary stabilized optical frequency combs. These pulses are exchanged between the nodes to simulate the Doppler shifts observed from a changing optical path length. We can continuously scan the simulated velocity from 14 to 620 m/s while simultaneously measuring velocity-dependent clock shifts at much higher velocities than has been previously recorded. This system provides an effective testbed that allows us to explore issues and solutions to enable femtosecond-level optical time transfer at high velocity.

## 1. Introduction

Precisely synchronized clocks enable many promising studies in relativity [1,2], cosmology [3-7], and geodesy [8-11]. Techniques that enable femtosecond optical two-way time-frequency transfer (O-TWTFT) have been well established over the last several years [12-21]. Giorgetta *et al*. at the National Institute of Standards and Technology (NIST) first demonstrated it was possible to conduct frequency comparisons using optical frequency combs and linear optical sampling [12]. Researchers at NIST were then able to demonstrate the synchronization of optical clocks to the sub-femtosecond level across a 4-km turbulent air path [13]. In addition, they showed that this level of synchronization was possible when the optical path between the nodes was increasing or decreasing at rates up to 25 m/s [14,15].

However, many applications require time transfer between nodes that are moving at much higher relative velocities. Above 50 m/s, many significant technical issues start to manifest such as coarse time transfer sub-system failure, interferogram (IGM) Nyquist edges, delay-Doppler coupling, and IGM rates going to zero [14]. These issues threaten to dramatically degrade the quality of time transfer at higher velocities jeopardizing applications that require femtosecond-level synchronization. Consequently, it is vital for future developments that O-TWTFT techniques are carefully tested and optimized for high-speed applications. This problem is further complicated by the difficulty in simulating high relative node velocities. A retroreflector placed on a quadcopter aircraft requires careful beam steering and tracking techniques [14,15], and the cost and complexity of operating faster aircraft makes systematic testing difficult. Multiple retroreflectors placed on a moving track in the lab are limited by optical alignment complexity, finite track lengths, and lab safety. Therefore, it is essential to develop a system that can systematically test O-TWTFT at a wide range of velocities that does not have these limitations.

In this work, we describe the development of a Doppler Simulator capable of simulating relative velocities over 600 m/s with no moving parts [22]. Instead of a computer model, the Doppler Simulator uses two optical clocks combined with optical frequency combs, optical fiber, modulators, and control electronics which synchronize the optical clocks in real time. This allows us to simulate a high Doppler-shift O-TWTFT system that will fully illuminate any issues that would occur in real motion. The system is highly versatile and complicated velocity profiles can be pre-loaded to simulate real-time clock synchronization in ground-to-aircraft, aircraft-to-aircraft, and satellite-to-satellite scenarios.

## 2. Experimental Setup

### *2.1 O-TWTFT with Changing Optical Path Length*

Our Doppler Simulator is an extended version of an O-TWTFT system based on optical frequency combs that can maintain femtosecond-level synchronization with a changing optical path length. This time transfer technique was developed by Sinclair *et al*. [14], and we show a simplified illustration in Fig. 1 (a). We built and tested a comparable system before we modified it to run as a Doppler Simulator. Note that this system does not account for true motion because neither of the nodes is physically moving. Instead, in both Ref. 14 and in our equivalent system, the individual nodes are stationary while the optical distance between them changes as the free-space portion of the beam is reflected off a moving retroreflector. As a result, special relativity can be ignored in both systems.

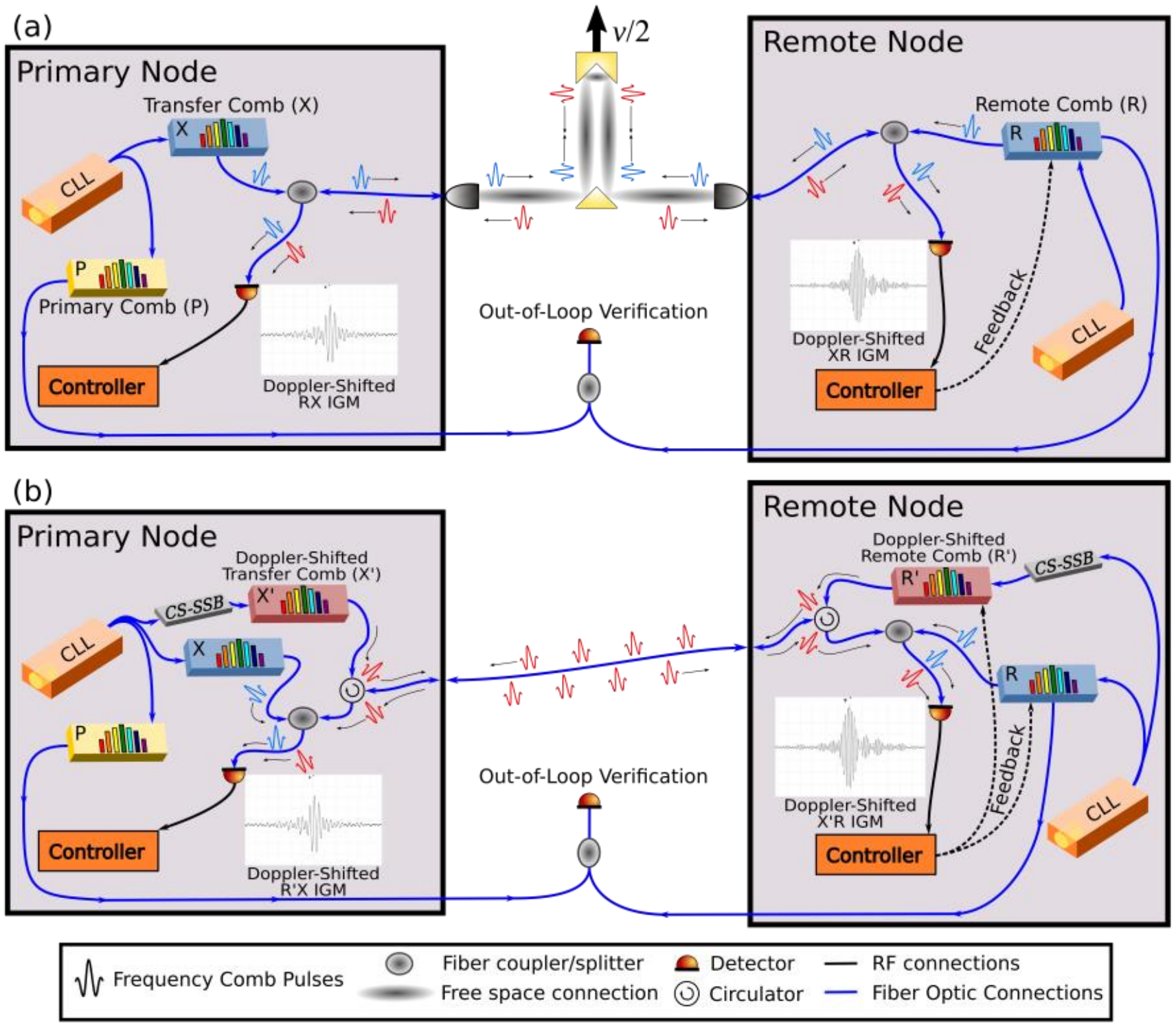

Fig. 1. (a) A femtosecond-level O-TWTFT system where a retroreflector moving at $v/2$ increases the optical path between nodes at a rate $v$ causing the exchanged comb pulses to experience a Doppler shift. (b) The Doppler simulator described in this work. The moving retroreflector is replaced with two separate combs (X' and R') which are locked to frequency-shifted cavity-locked laser (CLL) clock light using carrier suppressed single-sideband (CS-SSB) modulators to produce equivalent Doppler-shifted exchanged pulses.

Both the primary and remote nodes contain a narrow-line cavity-locked laser (CLL) from Stable Laser Systems (SLS-INT-1550-200-1) which serves as a stable optical reference. The CLL on the primary node is locked to both a primary comb (P) and a transfer comb (X) with nominal pulse repetition rates ($f_{rep}$) of 200 MHz. The two combs are locked ten mode numbers apart so that the transfer comb's pulse repetition rate difference ($\Delta f_{rep}$) is about 2 kHz. At the remote node, the remote comb (R) is also locked to its corresponding CLL at almost the same repetition rate as the primary comb. Transfer comb pulses are sent to the remote node and remote comb pulses are sent to the primary node where they interfere with the local combs producing an IGM at a rate of $1/\Delta f_{rep}$. The use of a transfer comb and the formation of IGMs, sometimes called linear optical sampling, are the key to enabling femtosecond resolution of the optical transmission time between nodes since observing the peak locations of the IGMs leverages the timing position of the underlying pulses by a factor of $f_{rep}/\Delta f_{rep} \sim 100{,}000$. In addition, a coarse timing link (not shown in Figure 1) exchanges timing information between nodes and carries pseudo-random binary sequences (PRBSs) to establish absolute round-trip timing information and overcome the 5 ns pulse-to-pulse ambiguity. One significant improvement we have made to our system is that we can operate the coarse timing link with either amplitude modulation (with on-off keying) or phase modulated communications [13]. Because the coarse timing link runs at only 10 Mb/s, the on-off keying communications are virtually immune to Doppler-related issues. As a result, we can conduct motion measurements at much higher velocities than those reported by Sinclair *et al.* [14,15]. The timestamps of the PRBSs and IGMs are recorded by the controller. The timestamps of the IGMs are used to refine the arrival times of the PRBSs at the primary ($T_{1,fine}$ and $T_{4,fine}$) and remote sites ($T_{2,fine}$ and $T_{3,fine}$). The clock difference between the primary and remote nodes where the optical path is changing by rate $v$, is given as

$$\Delta T_{P,R} = \frac{1}{2}\left(T_{1,fine} - T_{2,fine} - T_{3,fine} + T_{4,fine}\right) + \frac{v}{4c}\left(T_{1,fine} + T_{2,fine} - T_{3,fine} - T_{4,fine}\right) + \tau_{plane}, \quad (1)$$

where $\tau_{plane}$ is the relative offset between primary and remote combs which can be adjusted to overlap pulses at the out-of-loop verification reference plane [14]. The primary and remote combs have a 1 MHz difference in their carrier envelope offset lock point which produces a 1 MHz RF heterodyne signal when the comb pulses temporally overlap. With the right value of $\tau_{plane}$, the primary and remote clock difference is directly proportional to the amplitude of this heterodyne signal.

We use several Doppler compensation techniques to improve the optical time transfer performance of our system. We place dispersion compensation fiber to mitigate optical dispersion in the path. A matched filter is used on the primary-transfer IGM to improve timing measurements. The arrival time of the transfer-remote (XR) and remote-transfer (RX) IGMs is optimized using a cross-ambiguity function (CAF) which finds the effective arrival time of the Doppler shifted IGMs and compensates delay-Doppler coupling. Finally, we apply an additional matched filter on the XR and RX IGMs to compensate for RF dispersion in detectors, associated cables, and electronics.

### *2.2 High-Velocity Doppler Simulator*

To turn our O-TWTFT system into a Doppler Simulator (Fig 1(b)), we replaced the moving retroreflector with two optical frequency combs (X' and R') which are sent directly across the link. The Doppler shifted X' and R' comb pulses are routed with a circulator to cross the link and interfere with the unshifted R and X comb pulses and produce IGMs (X'R and R'X) that are identical to IGMs produced with a moving retroreflector. The X' and R' combs are "clones" of the transfer and remote combs, but the clock lasers that are used to lock these combs have been frequency shifted by a carrier-suppressed, single-sideband (CS-SSB) optical modulator before they mix with the comb light. Since the two clock lasers have slightly different wavelengths (1556.218 nm for the primary and 1556.603nm for the remote), the Doppler shift for each node is different and must be accurately calculated. Specifically, the shift is given by

$$f_{Dopp} = M f_{rep} \frac{v_{sim}}{c} \qquad (2)$$

where M is the mode number of the frequency comb tooth locked to the clock laser and $v_{sim}$ is the simulated velocity. For our system, $f_{rep}$ is 199.999492 MHz and the mode numbers are 963211 and 962973 for the primary and remote nodes, respectively. The Doppler shifts are generated by two AD9912 DDS boards with Teensy 4.0 microcontrollers clocked by a Rb stabilized 1 GHz RF generator and capable of millihertz resolution which are amplified and sent to the CS-SSB modulators. The maximum Doppler shift they can produce is ~400 MHz corresponding to a simulated velocity of ~620 m/s. The simulated velocity is changed via Python software running on a control computer. Intermediate values of the simulated velocity are calculated by the computer, and the full array of velocities are pre-loaded on to both DDS boards. Different acceleration profiles can be simulated by changing the step size of the pre-loaded velocity array. Once loaded, the primary node DDS board provides a trigger for the remote node DDS board so that the simulated velocity changes occur within ~50 ns for both boards ensuring the Doppler shift on both sides is consistent.

In addition to simulating Doppler shifts experienced by comb pulses crossing the optical link, we also adjust the timestamps generated by the coarse timing link to allow for the extra time that a PRBS would need to travel if the optical path length were also increasing. As is the case with a standard O-TWTFT system, the coarse time transfer algorithm is initiated when the primary node detects an IGM produced by remote pulses arriving and interfering with the local transfer comb. The primary node then transmits a PRBS at local time $T_{1,coarse}$ which is received at the remote node $T_{2,coarse}$ (local time). The remote node transmits its own PRBS at time $T_{3,coarse}$ which is received back at the primary node at time $T_{4,coarse}$. Because the physical optical path length does not change for our system, the last three values need to be modified for the coarse timing link to also simulate a changing path length. To do this, we introduced the recursively increasing values $\delta_1$ and $\delta_2$ which must account for the additional one-way optical path time from the primary to the remote and the remote to primary, respectively. This additional optical path time must also account for the Doppler shift of the PRBSs arrival rate at the remote and primary nodes. These values are given by

$$\delta_1 = \delta_{1,last} + \left(T_{2,coarse} - T_{2,coarse}^{last}\right)(v_{sim}/c)(1 + v_{sim}/c) \qquad (3)$$

$$\delta_2 = \delta_{2,last} + \left(T_{4,coarse} - T_{4,coarse}^{last}\right)(v_{sim}/c)(1 + 2v_{sim}/c) \qquad (4)$$

where the second term is the expected increase in optical time based on how far the remote node would have moved since the last timestamp was recorded. The factor of two in Eq. 4 accounts for the double Doppler shift experienced by PRBS rates traveling from the primary node to the remote node and back. Thus, during each processing loop we modify the Doppler-shifted coarse timestamps as follows:

$$T_{1,Dopp} = T_{1,coarse} \qquad (5)$$

$$T_{2,Dopp} = T_{2,coarse} + \delta_1 \qquad (6)$$

$$T_{3,Dopp} = T_{3,coarse} + \delta_1 \qquad (7)$$

$$T_{4,Dopp} = T_{4,coarse} + \delta_1 + \delta_2. \qquad (8)$$

The $T_{1,coarse}$ timestamp is unaltered. The timestamps $T_{2,coarse}$ and $T_{3,coarse}$ are increased by $\delta_1$ to allow for the simulated increasing time that it takes the primary PRBS to cross the link and the remote node to begin transmission of its PRBS. The timestamp $T_{4,coarse}$ is increased by $\delta_1$ and $\delta_2$ to allow for the increase in simulated full round-trip time. These timestamp modifications occur at every processing cycle when an R'X IGM is received at the primary node.

In addition to these modifications, we found that the system is very sensitive to thermal fluctuations. Optical pulses coming from the X' and R' combs have a substantial amount of nonreciprocal optical fiber that is not present in the standard system and thermal length changes in this fiber can cause the out-of-loop error signal to drift. As a result, we packaged the X' and R' combs along with all nonreciprocal fiber lengths inside a thermally isolated, temperature-controlled cabinet (Fig. 2).

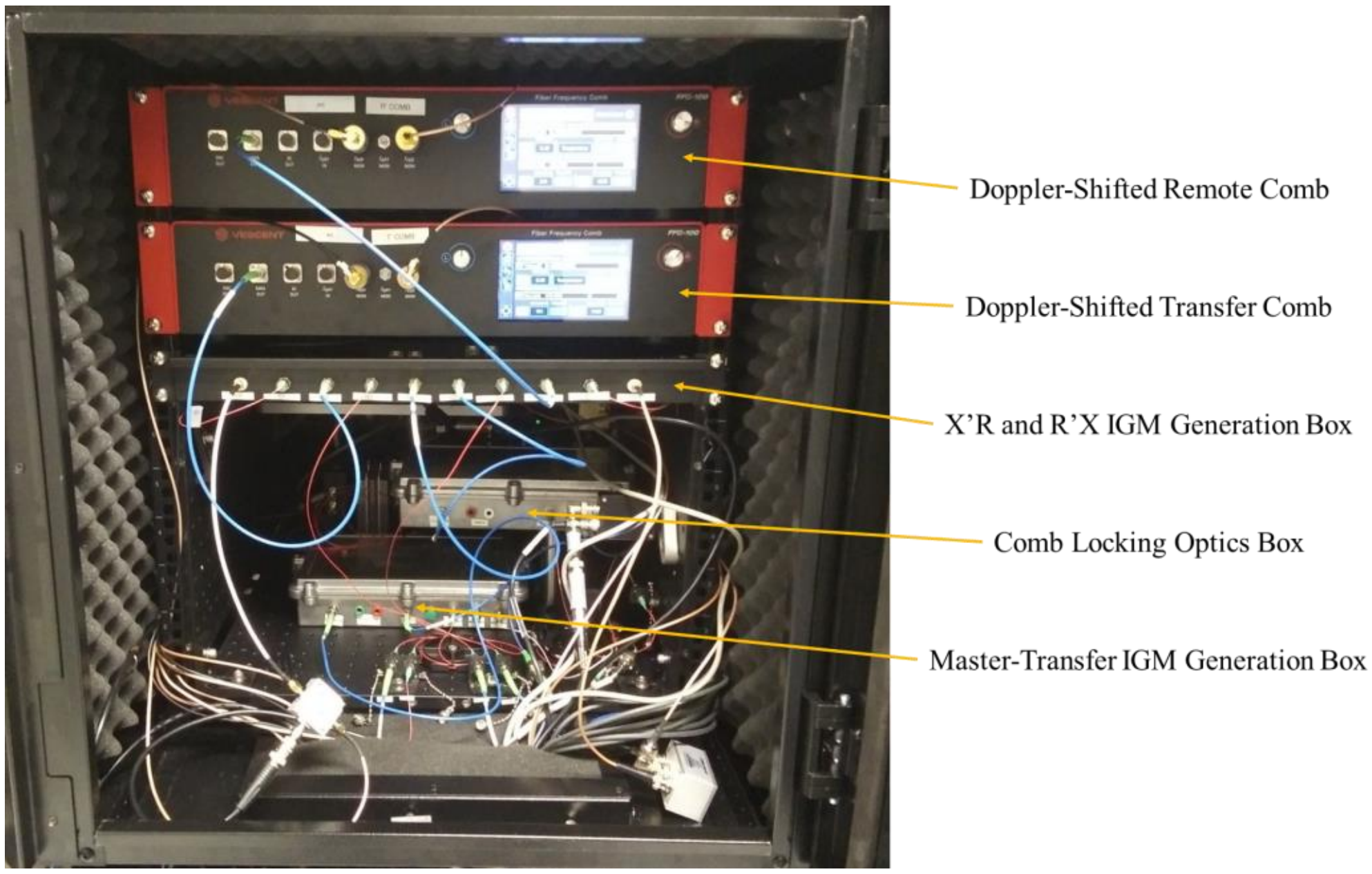

Fig 2. Thermally isolated, temperature-controlled cabinet containing all the temperature-sensitive nonreciprocal fiber within the Doppler Simulator system. The Doppler-shifted combs, IGM generation boxes and other locking optics are visible. Optical detectors and comb locking electronics are hidden behind.

Three 18"x12" water-cooled breadboards are mounted on the floor and sidewalls of the cabinet and kept at 10°C to keep the optical frequency combs from overheating. In addition, the cabinet contains all IGM generation optics, CS-SSB modulators and controllers, Red Pitaya comb controllers for the X' and R' combs, and all optical detectors. We found that the cabinet provided adequate thermal isolation to take velocity scans lasting several minutes with negligible thermal drift.

The Doppler Simulator can be scanned over hundreds of m/s with different accelerations. The minimum simulated velocity is 14 m/s (~9 MHz Doppler shift) and is limited by the bandwidth of the 90° hybrid RF couplers used to drive the CS-SSB modulators. As a result, velocity profiles that pass through zero are not possible, but it is possible to have negative velocity profiles (nodes approaching each other) if care is taken to ensure that the round-trip time does not become negative. As stated earlier, the maximum simulated velocity is 620 m/s (400 MHz Doppler shift) which is limited by the maximum frequency output of the AD9912 DDS boards which are clocked at 1 GHz. Accelerations are limited to about 225 m/s$^2$ (23*g*) because the Doppler-shifted combs are unable to maintain an optical lock on the frequency-shifted clock laser if the simulated velocity changes too quickly.

Even with all these modifications, we can quickly switch the system back to the standard O-TWTFT system shown in Fig 1(a). This allows us to quickly test upgrades on the system and investigate these changes with an actual changing path length.

## 3. Results

To validate the results that we obtained with the Doppler Simulator, we compared them with the clock synchronization error that we found by changing the optical distance between nodes via several large hollow corner cube retroreflectors on a moving optical rail system. Figure 3 shows a picture of the optical rail system with approximate beam locations indicated. The range of motion of the rail system is 192 cm. Beams from the primary and remote nodes are combined at a polarizing beam splitter and sent into the system. Each retroreflector reflection is spatially offset from the center of the reflector using 45° mirror mounts and smaller, stationary retroreflectors so that the beam reflects three times off each of the lower retroreflectors and twice off the upper reflectors. The beam is then sent back the way it came with a flat mirror so it again strikes the moving retroreflectors again. Once returned to the launch site, the beams are seperated by the polarizing beam splitter after passing through a quarter-wave/half-wave plate combination. The top speed of the retroreflectors on the rail is 3 m/s, but the multiple reflections off the retroreflectors increases the effective speed by a factor of 40 to 120 m/s.

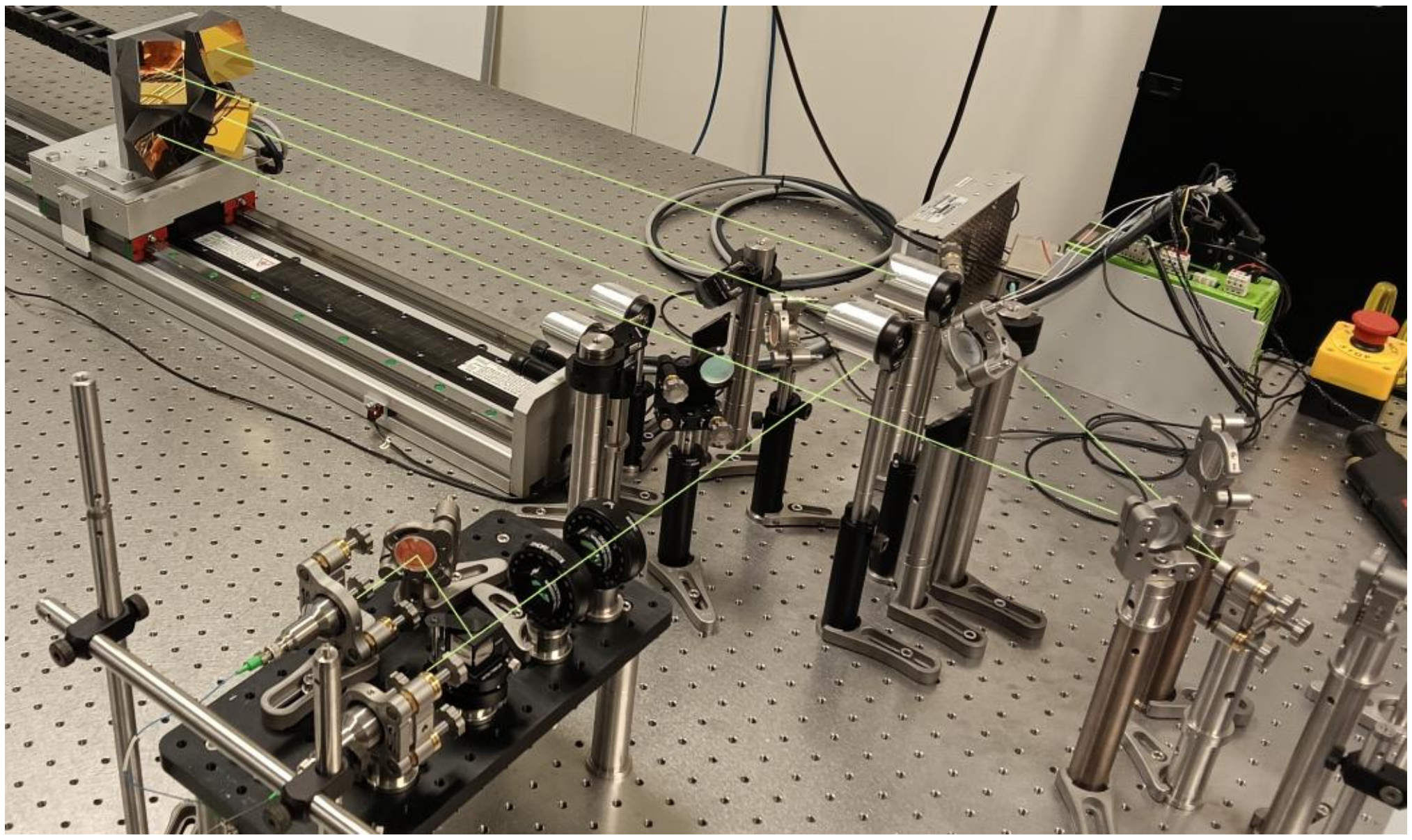

Fig. 3. The optical rail system we use to change the optical path length up to an effective speed of 120 m/s. The lines show rough beam locations; it would not be practical to show all beam locations. The beam is spatially offset from the center of the large moving hollow corner cube retroreflectors (visible on the left) so that it will reflect three times off the lower retroreflectors and twice off the upper retroreflectors.

In Figure 4, we compare the changing optical path length clock measurements with the results from the Doppler Simulator. The inset of Figure 4 shows an example how we calculate the clock changes from the optical rail system. After a 1-second stop at the end of the rail, we quickly accelerate the retroreflectors up to a fixed speed producing a large effective optical path change – 26 m/s in this case. While at this constant speed, we calculate the average out-of-loop clock difference and compare it with the average clock difference when the retroreflectors are stationary. The standard deviation of the clock difference is also calculated to produce error bars at each speed. This data was taken from 14 to 120 m/s at 2 m/s intervals.

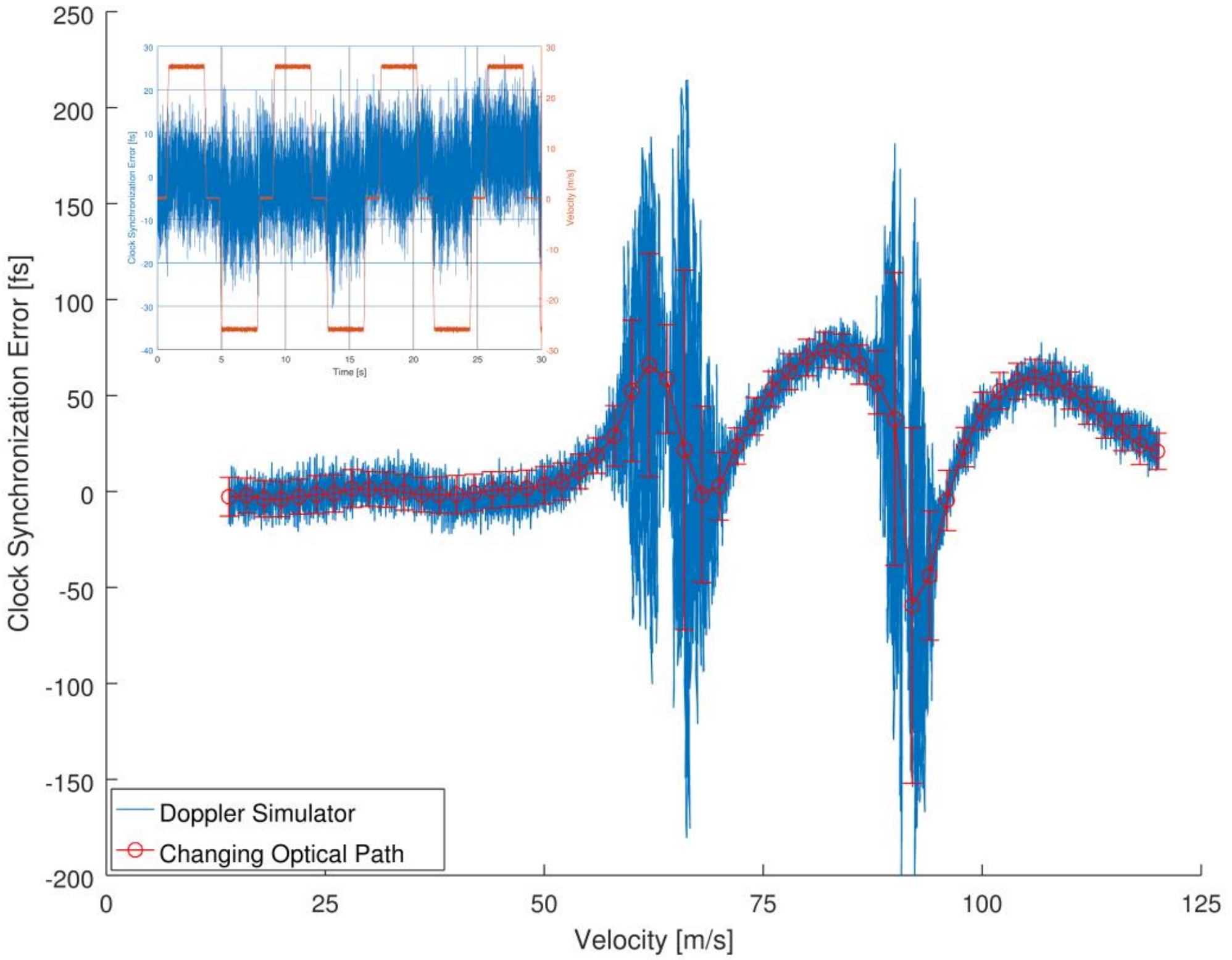

Fig. 4. The clock synchronization error observed with the Doppler Simulator and compared with the optical rail system with moving retroreflectors. The inset shows an example of the optical rail data at 26 m/s.

As can be seen, the agreement between the Doppler Simulator and the rail system is very good, giving us confidence that the Doppler Simulator results we observe at much higher velocities are accurate.

In Fig. 5, we show the clock synchronization error in a velocity scan with the Doppler Simulator from 14 to 620 m/s. This scan has a uniform acceleration of 5.1 m/s$^2$ lasting for about 120 s. We have done equivalent scans at 1.4 m/s$^2$ and 11.1 m/s$^2$ and did not observe any significant change in the clock synchronization error.

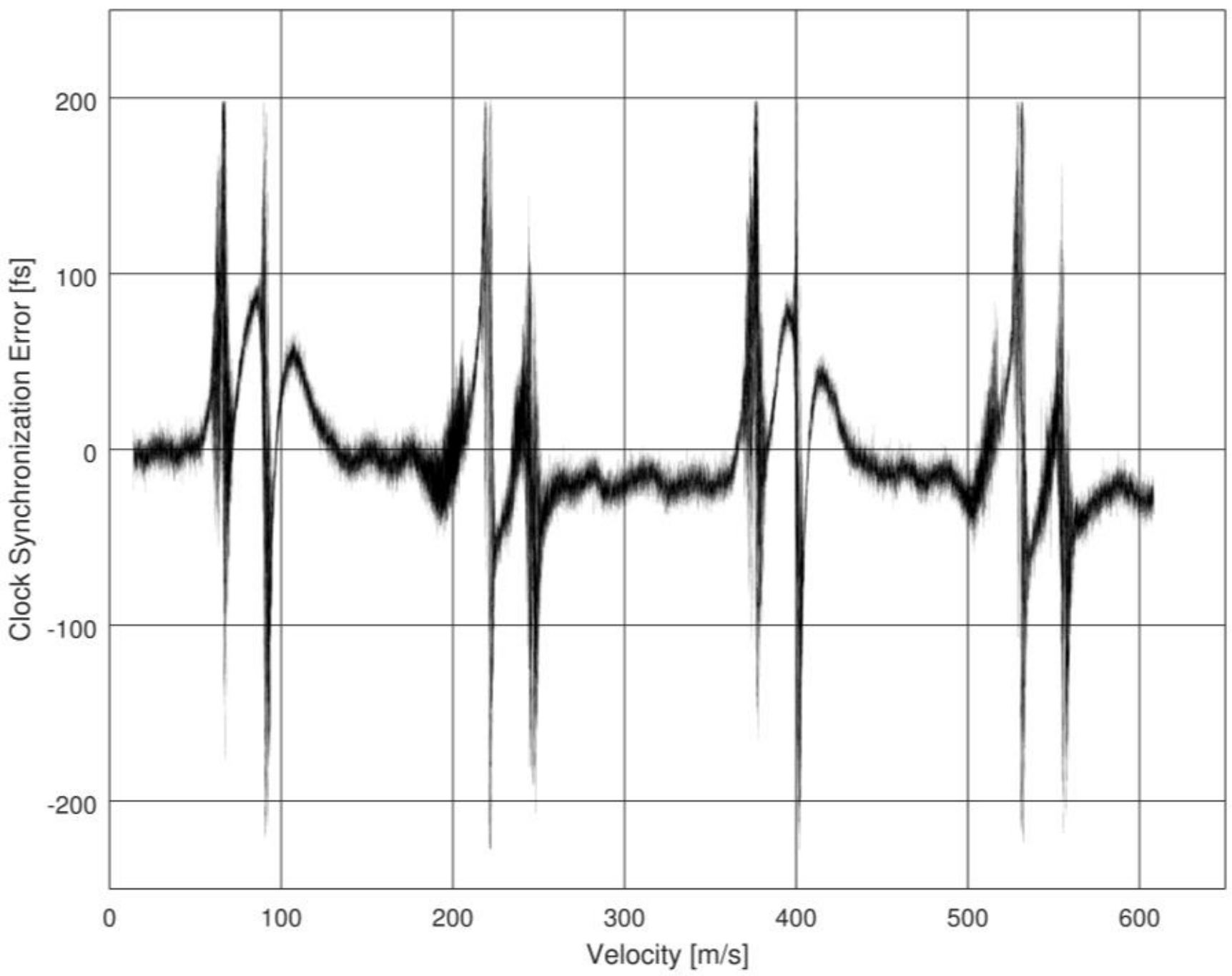

Fig. 5. The clock synchronization error observed with the Doppler Simulator while we scan the velocity from 14 to 620 m/s. In this scan we pass through four Nyquist zone edges.

The most obvious features are the four Nyquist zone edges which cause substantial clock errors and noise. At zero velocity, the transfer comb is nominally set so that the interfering comb spectral lines are approximately 50 MHz apart when centered on the optical filter placed before the photodetector (Fig. 6(b)). The resulting "carrier" frequency of the IGM is therefore 50 MHz. As the spectral lines are Doppler shifted with respect to one another, the interfering comb lines can reach a point where they are either overlapping (Fig. 6(a)) or half the repetition rate apart (~100 MHz, Fig 6(c)). In both cases, the IGM becomes very unstable as small phase differences between the interfering comb lines cause aliasing between generated beat notes. As a result, the shape of the IGM changes dramatically every time it's generated and recorded in the system. This effect can be clearly seen in oscilloscope traces of the generated IGMs in Fig. 6.

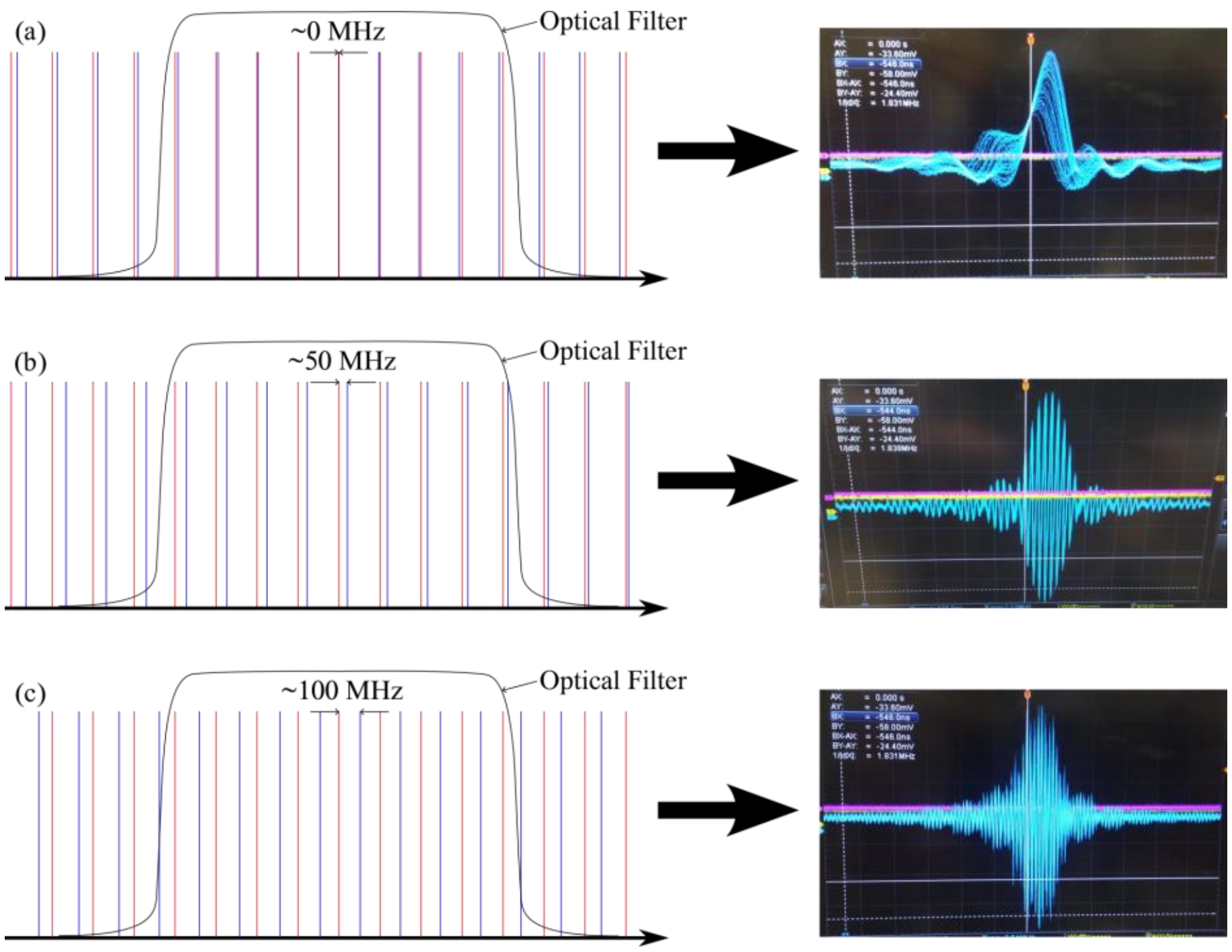

Fig. 6. The effect of aliasing in the interfering optical spectral lines between interfering combs at different Doppler shifts. The blue and red lines indicate different comb spectral lines within the optical filter. Nominally, the lines are 50 MHz apart (b) when the simulated velocity is zero. At different simulated velocities the IGM becomes unstable when the lines almost overlap (a) or are ~100 MHz apart (c).

As a result of this IGM instability, the Hilbert transform used to find the IGM peak fails to provide an accurate envelope of the IGM and the generated IGM timestamp will fluctuate significantly. This effect is an unavoidable consequence of the IGM generation process.

If the X'R and R'X IGMs both have a "carrier" frequency of 50 MHz while stationary, both will be unstable at 78 m/s corresponding to a Doppler shift of 50 MHz. Under those conditions, the system is unable to maintain a real-time time-frequency transfer lock when passing through that velocity. To mitigate this effect, we adjust the lock point for the X (and correspondingly the X') comb such that the X'R and R'X IGMs do not pass through their respective Nyquist edges at the same velocity. This widens the range of velocities where the clock error is influenced by the Nyquist edges and produces two regions of instability at each edge, but the severity of the instability is reduced to manageable levels. This lock point adjustment also results in the third Nyquist edge region to resemble the first because the IGMs go through the same type of deformation. The same is true of the second and fourth Nyquist edges.

The final observation we can make from Figure 5 is that the clock error has a ~30 fs linear drift from 14 to 620 m/s. We believe that this drift is caused by CAF filters insufficiently compensating for the delay-Doppler coupling at higher simulated velocities. This is because we have observed a change in the linear drift with additional dispersive fiber in the system. Changing the amount of dispersion in the system will also modestly affect the shape of the curve around the Nyquist zone edges.

We also investigated the effect of acceleration on the O-TWTFT protocol. Our results are summarized in Figure 6, which shows the result of velocity scans from 14 to 50 m/s with accelerations ranging from 0.43 m/s$^2$ (0.044$g$) to 226 m/s$^2$ (23$g$).

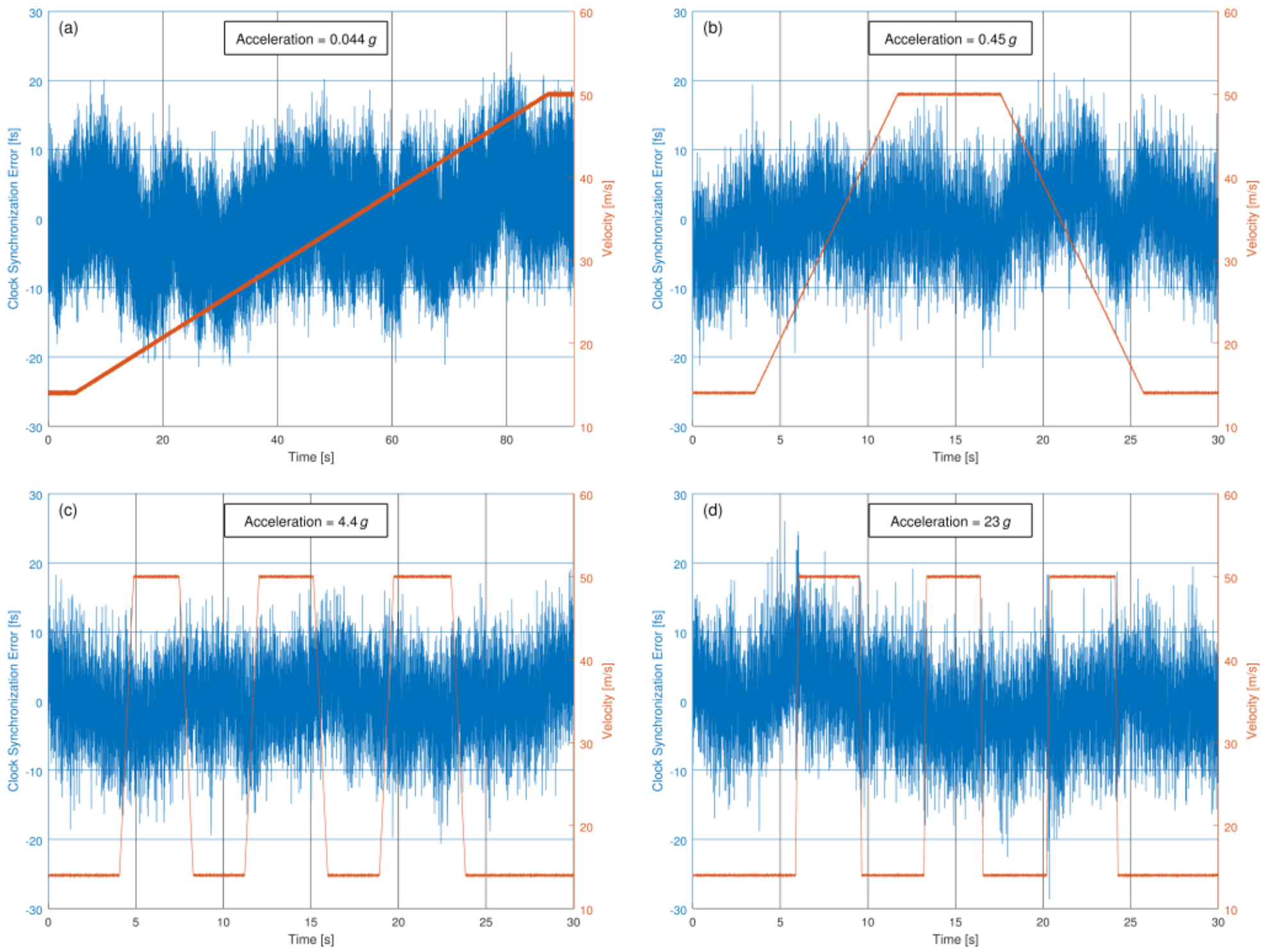

Fig. 7. The clock synchronization error measured with the Doppler Simulator when the system is scanned between 14 and 50 m/s with an acceleration of (a) 0.044*g*, (b) 0.45*g*, (c) 4.4*g*, and (d) 23*g*.

We found that acceleration does not significantly affect the clock error within this velocity range. Small spikes in the clock error start to appear in data at 23*g*, but these spikes are insignificant compared to the background noise. In addition, optical frequency combs and associated electronics are unlikely to survive prolonged exposure to 23*g*'s of real acceleration.

## 4. Discussion and Future Work

We have shown that the Doppler Simulator is capable of testing real-time O-TWTFT systems at velocities far beyond what can be done in a laboratory environment, and we have provided evidence that the Simulator produces accurate real-world results. Our early testing of the system indicates the presence of Nyquist edges that cause a significant reduction in the performance of a time transfer system at certain velocities. In addition, the Doppler Simulator indicates that this O-TWTFT system is virtually immune to acceleration-related clock errors.

Moving forward, we are developing strategies that will further mitigate the effects of the Nyquist edges such as careful error filtering to reduce the influence of Hilbert transform errors at certain velocities. Since we have complete control of the Doppler shift imparted to the combs, we believe that we can modify the Doppler Simulator to simulate special and general relativity. We are also planning to modify the Doppler Simulator to make it capable of simulating even higher velocities (>1 km/s) by replacing the DDS boards with ones that can produce higher frequencies so we can produce greater Doppler shifts. In this way we will improve the Doppler Simulator into an even more effective tool for mitigating Doppler and relativity-related effects in O-TWTFT systems.

**Funding.**

This work is supported by the Air Force Research Laboratory.

### Disclosures.

The authors declare no conflicts of interest.

### Acknowledgements.

Approved for public release, distribution is unlimited. Public Affairs release approval #AFRL-2025-2375. The views expressed are those of the authors and do not reflect the official policy or position of the Department of the Air Force, the Department of Defense, or the U.S. government.

### Data Availability Statement (DAS).

Data underlying the results presented in this paper are not publicly available at this time but may be obtained from the authors upon reasonable request.

### References

1. C.W. Chou, D.B. Hume, T. Rosenband *et al.*, "Optical clocks and relativity," Science **329,** 1630–1633, (2010).
2. M. Takamoto, I. Ushijima, N. Ohmae *et al.*, "Test of general relativity by a pair of transportable optical lattice clocks," Nat. Photon. **14,** 411–415 (2020).
3. Y.-D. Tsai, J. Eby, and M.S. Safronova, "Direct detection of ultralight dark matter bound to the Sun with space quantum sensors," Nat. Astron. **7,** 113–121 (2023).
4. A. Arvanitaki, J.W. Huang, and K. Van Tilburg, "Searching for dilaton dark matter with atomic clocks," Phys. Rev. D **91,** 015015 (2015).
5. A. Derevianko and M. Pospelov, "Hunting for topological dark matter with atomic clocks," Nat. Phys. **10,** 933–936, (2014).
6. P. Wcisło, P. Ablewski, K. Beloy, *et al*., "New bounds on dark matter coupling from a global network of optical atomic clocks," Sci. Adv. **4,** eaau4869, (2018).
7. C.J. Kennedy, E. Oelker, J.M. Robinson, *et al*., "Precision metrology meets cosmology: Improved constraints on ultralight dark matter from atom-cavity frequency comparisons," Phys. Rev. Lett. **125,** 201302, (2020).
8. T. Takano, M. Takamoto, I. Ushijima, *et al.*, "Geopotential measurements with synchronously linked optical lattice clocks," Nat. Photon. **10,** 662–666 (2016).
9. S. Kolkowitz, I. Pikovski, N. Langellier, *et al.*, "Gravitational wave detection with optical lattice atomic clocks," Phys. Rev. D **94,** 124043 (2016).
10. W.F. McGrew, X. Zhang, R. J. Fasano, *et al.*, "Atomic clock performance enabling geodesy below the centimetre level," Nature **564,** 87–90, (2018).
11. J. Grotti, S. Koller, S. Vogt, et al., "Geodesy and metrology with a transportable optical clock," Nat. Phys. **14,** 437–441, (2018).
12. F.R. Giorgetta, W.C. Swan, L.C. Sinclair, *et al.*, "Optical two-way time and frequency transfer over free space," Nat. Photonics **7,** 434 (2013).
13. J.-D. Deschênes, L.C. Sinclair, F.R. Giorgetta, *et al.*, "Synchronization of distant optical clocks at the femtosecond level," Phys. Rev. X **6,** 021016, (2016).
14. L.C. Sinclair, H. Bergeron, W.C. Swan, *et al.*, "Femtosecond optical two-way time-frequency transfer in the presence of motion," Phys. Rev. A **99,** 023844 (2019).
15. H. Bergeron, L.C. Sinclair, W.C. Swan, *et al.*, "Femtosecond time synchronization of optical clocks off of a flying quadcopter," Nature Communications **10,** 1819 (2019).
16. M.I. Bodine, J.-D. Deschênes, I. Khader, *et al.*, "Optical atomic clock comparison through turbulent air," Phys. Rev. Research **2,** 033395 (2020).
17. M.I. Bodine, J.L. Ellis, W.C. Swan, *et al.*, "Optical time-frequency transfer across a free-space, three-node network," APL Photonics **5,** 076113, (2020).
18. J.L. Ellis, M.I. Bodine, W.C. Swann, *et al.*, "Scaling up frequency-comb-based optical time transfer to long terrestrial distances," Phys. Rev. Applied **15,** 034002 (2021).
19. Boulder Atomic Clock Optical Network (BACON) Collaboration, "Frequency ratio measurements at 18-digit accuracy using an optical clock network," Nature **591,** 564-569 (2021).
20. E.D. Caldwell, J.-D. Deschênes, J.L. Ellis, *et al.*, "Quantum-limited optical time transfer for future geosynchronous links," Nature **618,** 721-726 (2023).
21. K.W. Martin, N. Zaki, N. Matthews, *et al.*, "Demonstrations of real-time precision optical time synchronization in a true three-node architecture," J. Phys. Photonics **7,** 025014 (2025).
22. M.S. Bigelow, K.W. Martin, J.D. Elgin, *et al.*, "Development of a high-doppler shift femtosecond optical time transfer testbed," in Conference on Lasers and Electro-Optics (CLEO) (2024) paper SF3D.4.